\documentclass[twocolumn,amsmath,amssymb]{revtex4}
\pdfoutput=1
\usepackage{xspace}
\usepackage{graphicx}
\usepackage{color}
\usepackage{slashed}
\def\beq{\begin{eqnarray}}
\def\eeq{\end{eqnarray}}

\newcommand{\roughly}[1]%
    {{\mathrel{\raise.3ex\hbox{$#1$\kern-.75em\lower1ex\hbox{$\sim$}}}}}

\newcommand{\Eq}[1]{Eq.~(\ref{#1})}

\begin{document}

%\preprint{APS/123-QED}

%\title{The Higgs as Harbinger of Non-Minimal Supersymmetry}
\title{Early Higgs Hints for Non-Minimal Supersymmetry}

\author{Aleksandr Azatov$^*$,\ \
Spencer Chang$^\dagger$,\ \  
Nathaniel Craig$^\ddagger$,\ \ 
Jamison Galloway$^*$\ \ 
}

\address{$^*$Dipartimento di Fisica, Universit\`a di Roma ``La Sapienza"\\
{\rm and} INFN Sezione di Roma, I-00185 Rome}

\address{$^\dagger$Department of Physics, University of Oregon, Eugene, OR 97403}

\address{$^\ddagger$Department of Physics, Rutgers University Piscataway, NJ 08854\\
{\rm and} School of Natural Sciences, Institute for Advanced Study 
Princeton, NJ 08540}

\begin{abstract}
\noindent
We discuss the role that Higgs coupling measurements can play in differentiating supersymmetric extensions of the Standard Model.  Fitting current LHC data to the Higgs couplings, we find that the likelihood fit shows a preference in the direction of suppressed (enhanced) bottom (top) quark couplings.  In the minimal supersymmetric Standard Model, we demonstrate that for $\tan \beta > 1$, there is tension in achieving such fermion couplings due to the structure of the Higgs quartic couplings.   In anticipation of interpreting supersymmetric models with future data, we determine a single straightforward condition required to access the region of coupling space preferred by current data.

\end{abstract}

%\pacs{12.60.Nz}% PACS, the Physics and Astronomy
                             % Classification Scheme.

\maketitle
\section{\label{sec:intro}Introduction}

The LHC is poised to accurately determine the mechanism of electroweak symmetry breaking, building upon the recent discovery of a new Higgs-like state near 125 GeV \cite{ATLAS,CMS}. 
Should this new state prove to be an elementary scalar, 
supersymmetry (SUSY) remains the principal candidate for stabilizing the electroweak hierarchy. However, the minimal supersymmetric Standard Model (MSSM) is somewhat strained to explain a Higgs at 125 GeV, requiring significant enhancement of the tree-level Higgs mass that is in tension with naturalness.  Here, we emphasize that 
%apart from naturalness considerations, 
the structure of the MSSM also tightly constrains the possible tree-level couplings of the Higgs. If the production and decay modes of the Higgs deviate from Standard Model predictions, it would not only be an indication of new physics, but may also decisively favor or disfavor the MSSM well before other states are discovered. 
To this end,
 the measurement of Higgs couplings provide a sensitive and immediate probe of physics above the weak scale.

In this work we perform a model-independent fit of Higgs couplings using current LHC data, focusing on implications for theories with two Higgs doublets. %such as the MSSM. 
%At present the statistical errors in Higgs search channels are large, but there is sufficient data to perform a tentative model-independent fit of Higgs couplings and begin exploring the implications for new physics.   
We find that the MSSM is facing tension with certain elements of the data.  
%At issue is the structure of its quartic Higgs potential, and a generic preference for enhanced coupling to down-type fermions.  
At issue is the structure of its quartic Higgs potential, leading to a generic preference for enhanced coupling to down-type fermions.
Indeed, the tree-level potential mandates such enhancement whenever $\tan \beta >1$ and we find even at loop-level that achieving significant suppression is atypical.  By analyzing the quartic terms in full generality, we show that this conclusion can be avoided and pinpoint parameter space for the MSSM and simple alternatives to accommodate suppressed couplings to down-type fermions.  
%In the event that future data support the values of couplings observed thus far, models involving  new non-minimal dynamics will stand out as the clear candidates for explaining weak-scale physics.

\section{\label{sec:Status}Status of Higgs Measurements}
\begin{figure}[htb]
\begin{center}
\includegraphics[width=7cm]{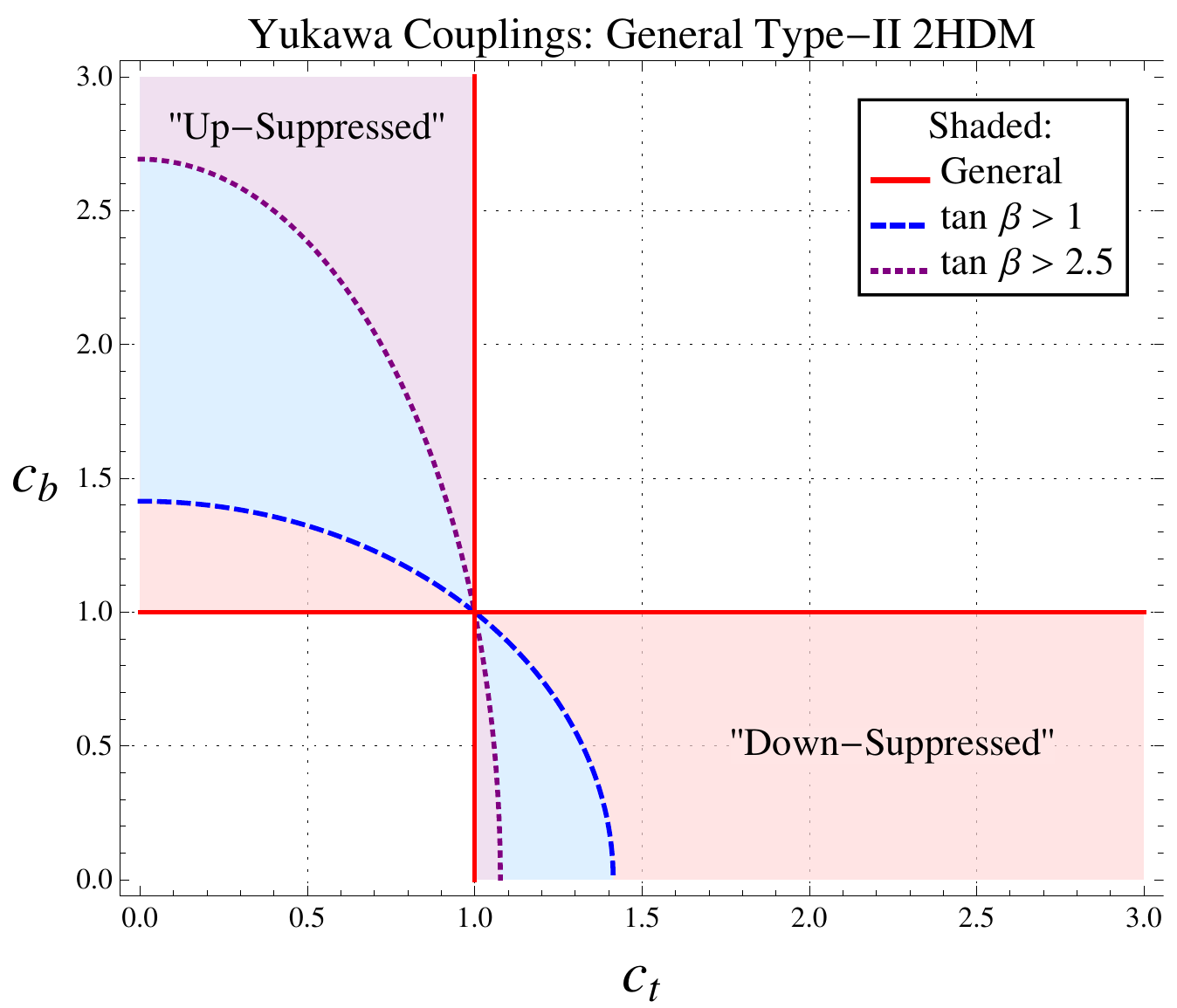}
\caption{The two regions accessible in a generic type-II 2HDM.  Down-type couplings are enhanced when up-type are suppressed and vice versa.  For the MSSM and simple extensions, the lower region is largely inaccessible when $\tan \beta > 1$.}
\label{fig:TB}
\end{center}
\end{figure}

We begin by establishing the relevant conventions for a type-II two Higgs doublet model (2HDM) like the MSSM.
%, where one Higgs couples only to up-type fermions while the other couples only to down-type. 
The mass eigenstates of the neutral CP-even states are
\beq
\begin{pmatrix}
h^0 \cr H^0 \end{pmatrix} =
\sqrt 2
\begin{pmatrix} -\sin \alpha & \cos \alpha \cr \cos \alpha & \sin \alpha \end{pmatrix}
\begin{pmatrix} {\rm Re}\, H_d^0 \cr {\rm Re} \, H_u^0 \end{pmatrix},
\eeq
with mixing angle $\alpha \in [-\pi/2,\pi/2]$.  The couplings of  the light eigenstate $h^0$ to SM fields are then given by
\beq\label{eq:a}
\hspace{0.5cm}a \equiv \frac{g_{hVV}}{g_{hVV}^{\rm SM}} = \sin(\beta -\alpha),\hspace{1.2cm} \\
\label{eq:cucd}
c_t \equiv \frac{g_{ht\bar t}}{g_{h t \bar t}^{\rm SM}} = \frac{\cos \alpha}{\sin \beta}, \quad
c_b \equiv \frac{g_{hb \bar b}}{g_{hb \bar b}^{\rm SM}} = -\frac{\sin \alpha}{\cos \beta},
\eeq
%\beq\label{eq:a}
%a &\equiv& \frac{g_{hVV}}{g_{hVV}^{\rm SM}} = \sin(\beta -\alpha), \\
%\label{eq:cu}
%c_u &\equiv& \frac{g_{hu\bar u}}{g_{h u \bar u}^{\rm SM}} = \frac{\cos \alpha}{\sin \beta}, \\
%\label{eq:cd}
%c_d &\equiv& \frac{g_{hd \bar d}}{g_{hd \bar d}^{\rm SM}} = -\frac{\sin \alpha}{\cos \beta};
%\eeq
%with charged leptons coupling like the down-type quarks; 
which we will refer to as the gauge coupling, and the up and down-type Yukawa couplings, respectively.  A full discussion can be found for instance in \cite{HiggsHunters}.  Thus the 2HDM has access to two distinct regions in the positive quadrant of 
%the plane spanned by up- and down-type 
Yukawa couplings, as illustrated in Fig.~\ref{fig:TB}.  %Since these couplings satisfy $\sin^2 \beta\, c_t^2 + \cos^2 \beta\, c_b^2 =1$, the 2HDM is constrained to two distinct regions where the couplings are i) top suppressed ($c_t < 1, c_b > 1$) or ii) bottom suppressed (($c_t > 1, c_b < 1$)). 
%We will see below how these regions stand up to what has been observed thus far at the LHC. \\
%%%%%%%%%%%%%%%%%%%%%%%%%%%%%
%%%%%%%%%%%%%%%%%%%%%%%%%%%%%
%%%%%%%%%%%%%%%%%%%%%%%%%%%%%
%%%%%%%%%%%%%%%%%%%%%%%%%%%%%

\begin{figure}[htb]
\begin{center}
\includegraphics[width=8cm]{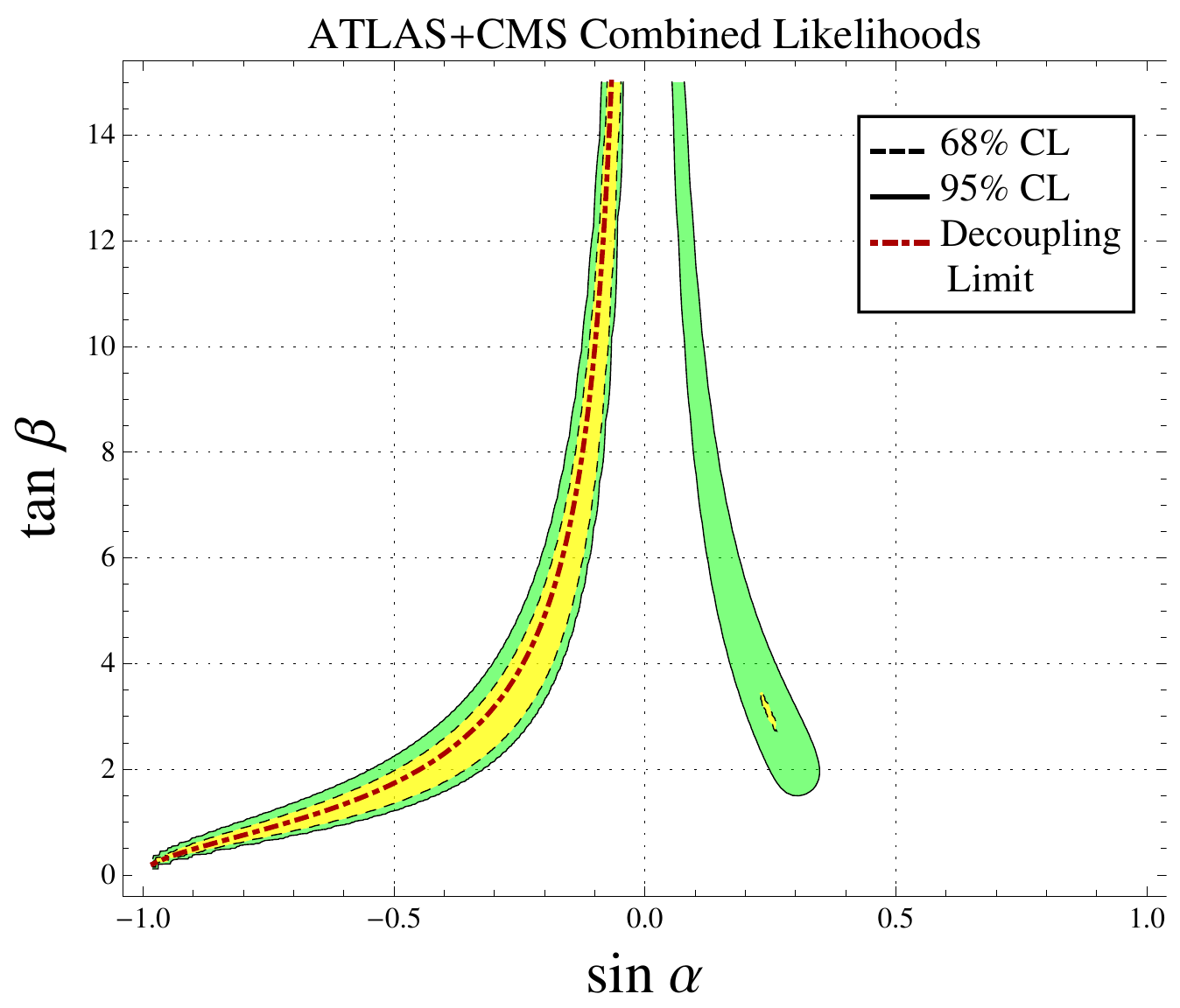}
\caption{Likelihoods drawn from Higgs searches at ATLAS and  CMS assuming that new loop-induced decays to SM states are small. The likelihood peaks below the decoupling contour, where down-type Yukawa couplings are suppressed.}
\label{fig:alphabeta}
\end{center}
\end{figure}

We now discuss the current experimental status of these Higgs couplings, which we will show can have an important impact on various SUSY scenarios.  As has been noted in recent literature, the likelihoods of the hinted state near 125 GeV viewed in the space spanned by $(\sin \alpha,\tan \beta)$ are concentrated near the decoupling limit (cf. \cite{Falkowski}) where $\alpha \to \beta -\pi/2$ and all couplings take their SM values.  Using current ATLAS \cite {ATLAS} and CMS \cite{CMS} results
% which provide data in fully exclusive modes
we demonstrate this in Fig.~\ref{fig:alphabeta} using exclusive best fit information, following the statistical method of
%the method for likelihood reconstruction detailed in 
\cite{exclusions} and assuming that loop-induced decays are dominated by their contributions from SM fields (as will be done throughout); alternative statistical methods yield consistent results \cite{Falkowski,fits}. 
%%%%%%%%%%%%%%%%%%%%%%%%%%%%%
%%%%%%%%%%%%%%%%%%%%%%%%%%%%%
%%%%%%%%%%%%%%%%%%%%%%%%%%%%%
%%%%%%%%%%%%%%%%%%%%%%%%%%%%%
We note, however, that slight deviations from the decoupling limit contour of Fig.~\ref{fig:alphabeta} can amount to significant changes in the Yukawas; for instance the lower limit of the 68\% CL contour corresponds to $c_b \simeq 0.7$ for $\tan \beta \gtrsim 2$.  Furthermore, flat priors in this space give  inequitable treatment of $v_u$ and $v_d$. Examining the space of the Yukawas is therefore important in its own right. 

\begin{figure}[htb]
\begin{center}
\includegraphics[width=8cm]{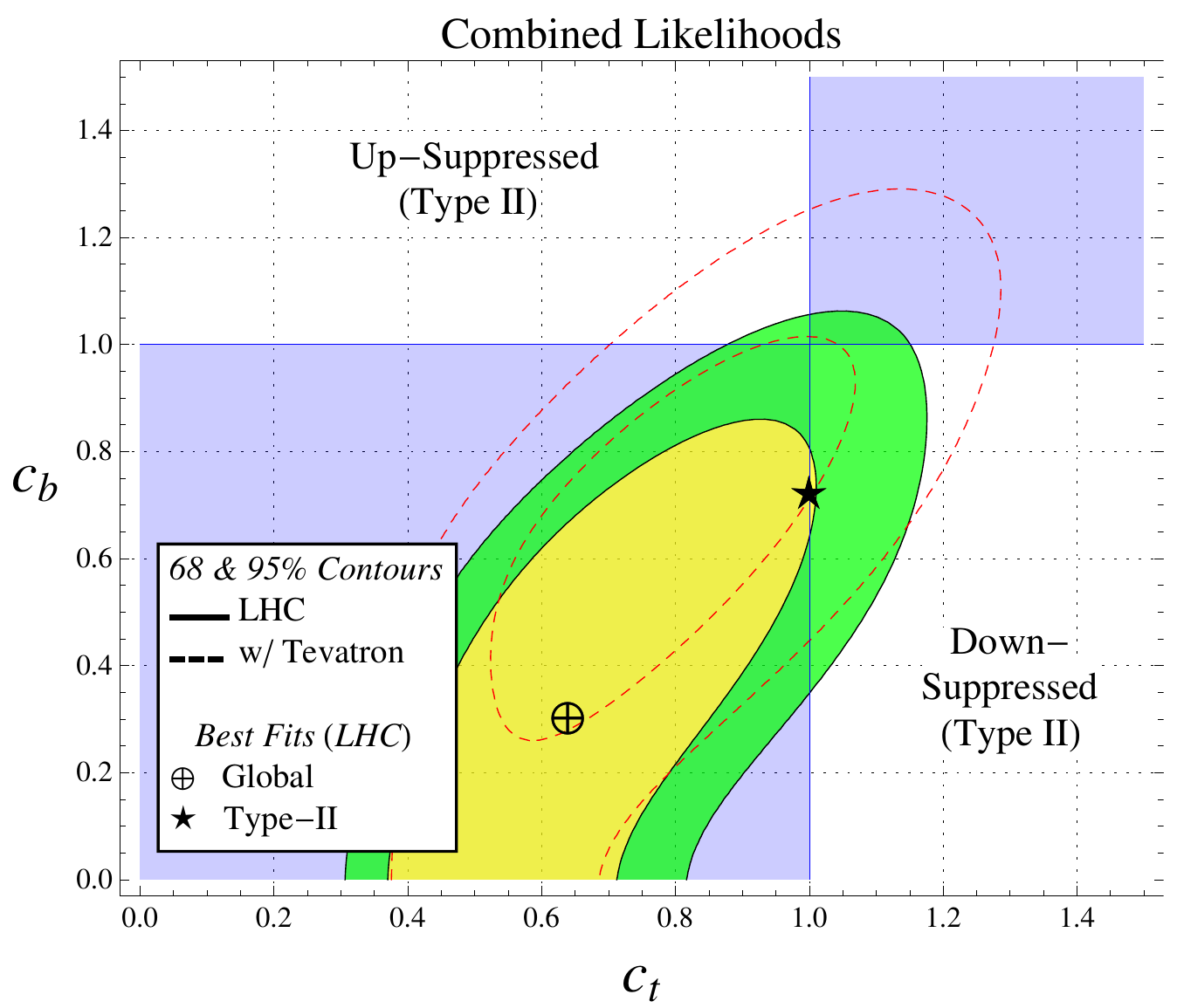}
\caption{Likelihoods constructed in the three-dimensional space of gauge and Yukawa couplings of the Higgs, marginalizing over the gauge coupling in the range $0 \leq a \leq 1$.   The colored contours show the space preferred by LHC data and the dashed red lines show how the fit is affected by the inclusion of Tevatron data.  The two unshaded regions are those accessible in a type-II 2HDM; for $\tan \beta>1$, the MSSM is typically constrained to the up-suppressed region (cf. Fig~\ref{fig:TB}).}
%top and bottom Yukawas currently prefer region II which would require $\tan \beta < 1$.}
\label{fig:3D}
\end{center}
\end{figure}

Analyzing the preferences of the Yukawa couplings can be done by constructing  likelihoods in the space $(a,\, c_b,\, c_t)$ and marginalizing over the vector coupling, $a$; we show these results in Fig.~\ref{fig:3D}.   The likelihood in this space has significant support only when $a\gtrsim 0.6$, and in this range we find that the best fit values for the fermion couplings always occur with $c_t$ near its SM value and $c_b$ suppressed as shown in Table~\ref{table:a_regions}; such a preference is further supported by the fits to fermion couplings recently reported by the ATLAS collaboration \cite{ATLAScouplings}.
Incidentally, the preference for suppressed down-type couplings is maintained even if the combined fit includes Tevatron data, which has shown indications of increased rates in $h \to b \bar b$ \cite{tevatron}.  The best fit for $c_b$ in the type-II space increases by $\mathcal O(10\%)$ with the inclusion of the Tevatron data, but still prefers values $<1$ as indicated in Fig.~\ref{fig:3D}.  
% though has only a minor impact on the global analysis due to limited statistical significance.  
As we will discuss in the next section, this preference for suppression of down-type couplings causes tension in the MSSM, which preferentially populates the up-suppressed region when $\tan \beta > 1$.   
\begin{table}[htb]
\begin{center}
\begin{tabular}{| c | c |}\hline
$a$ values &  $(\hat c_t, \hat c_b)$ \\ \hline
$0\leq a<0.2$ &  $(2.4,0.17)$ \\ 
$0.2\leq a<0.4$ &  $(1.0,0.1)$\\ 
$0.4\leq a<0.6$ & $(1.0,0.44)$ \\ 
$0.6\leq a<0.8$ &  $(1.0,0.62)$ \\ 
$0.8\leq a \leq 1$ &  $(1.0,0.89)$ \\ 
\hline
\end{tabular}
\caption{\small
Best fit points $(\hat c_t, \hat c_b)$ in the type-II parameter space for different slices of the vector coupling, $a$.  For all values of this coupling, the down-type Yukawa are preferentially suppressed.}
\label{table:a_regions}
\end{center}
\label{default}
\end{table}

%The reason for the tendency towards a decoupling limit should in any case be examined somewhat carefully.  For instance, the likelihood can be examined in the space spanned by $(a,\, c_b,\, c_t)$, in which case the picture becomes clearer.  

%The preferences for the fermion couplings can be more clearly understood by examining the likelihood in the couplings $(a,\, c_b,\, c_t)$.  In Fig.~\ref{fig:3D}, the preference for the decoupling limit is seen to arise due to a preference for SM-like gauge coupling, with likelihoods at $a\sim 1$ greater by a factor $\simeq 5$ than those at $a<0.5$.  Such a  preference is dominated by the significant excess that is currently reported in the $\gamma \gamma$ final state \cite{CMSgaga}.  This, however,  is at odds with the $WW/ZZ$ channels which prefer suppression compared to SM.    Because the combined data prefer the gauge coupling to be as large as possible, the decoupling limit is however optimal unless additional contributions to the $h \to \gamma \gamma$ width are included.  

%%%%%%%%%%%%%%%%%%%%%%%%%%%%%
%%%%%%%%%%%%%%%%%%%%%%%%%%%%%
%%%%%%%%%%%%%%%%%%%%%%%%%%%%%
%%%%%%%%%%%%%%%%%%%%%%%%%%%%%

%Also worth noting in Fig.~\ref{fig:3D} is that the likelihood peaks near the SM point, but that there exists a shallow direction into the down-suppressed region of Fig.~\ref{fig:TB}.  This stems from the fact that the model-independent best fit point in $\gamma \gamma$ occurs at $a=1.35,\, c_b=1.1,\, c_t=0.65$ while the likelihoods for the $VV$ channels peak near $a=0.4$  with top couplings substantially enhanced.  The latter is easily understood noting that the $WW+2j$ channel---that most sensitive to $a$---observes a large deficit, while the remaining $VV$ channels are essentially SM-like; accommodating these remaining $VV$ channels thus requires $c_t>1$ in order to enhance their (inclusive) production and compensate for reduced branching.  These points are further detailed by the population counts and best fit points shown in Table~\ref{table:a_regions}. 
%From this we conclude that if there are observable deviations in the fermion couplings from their SM values, the data are now hinting that it will be in the down suppressed region.  As we will discuss in the next section, this causes tension in the MSSM, which preferentially populates the up-suppressed region when $\tan \beta > 1$.   

%\section{\label{sec:Couplings}Fermion Couplings in the MSSM and General Type-II 2HDM}
\section{\label{sec:Couplings} General Type-II 2HDM Analysis}
We now consider the fermion couplings that occur in the general type-II 2HDM, identifying the conditions to have down-suppression.  For the MSSM at tree level, we will find that it is impossible to have down-suppression for $\tan \beta > 1$ and gain insight on why at loop level, down-suppression is pushed to a specific region of supersymmetry breaking parameter space.    

Before proceeding, we pause to remark on the strong evidence from ancillary considerations that $\tan \beta > 1$ in the MSSM. Simple perturbativity of Higgs-top interactions requires $\tan \beta \gtrsim 0.3$, while more conservatively avoiding Landau poles in the top yukawa below the unification scale imposes $\tan \beta \gtrsim 1.5.$ But the Higgs mass itself may provide the strongest lower bound on $\tan \beta$ in the MSSM; even provided a favorable tuning of stop masses and $A$-terms, the MSSM requires $\tan \beta \gtrsim 5$ to generate a Higgs mass near 125 GeV without additional contributions from new degrees of freedom \cite{tanbeta}.  Taking $\tan \beta <1$ to be essentially excluded, we can focus on the 2HDM dynamics that is required to achieve down-suppression for $\tan \beta > 1$.  
%the conclusion is that the up-suppressed region is the remaining space in this plane for such models at tree-level.  We will take up the issue of how this statement might be modified at loop-level, or by the addition of new dynamics, below. \\

\subsection{General Quartic Structure}

The possibility of increasing parametric freedom in the MSSM by coupling SUSY Higgses to new fields has been studied extensively. Such thinking is certainly attractive with respect to the mass of the lightest Higgs, as the hinted state at 125 GeV would require significant tuning within the MSSM.  Given the fits to LHC data presented above, it is equally useful to consider how such new parametric freedom could alter MSSM fermion couplings as a function of $\tan \beta$. 

We first consider the general structure of a type-II 2HDM, and determine the conditions under which down-type Yukawa couplings might be significantly suppressed at large $\tan \beta$.   We begin with the generic quartic terms that will contribute to the potential for the neutral fields:
% Choosing a convention which is well-suited for this---differing notably from more standard parameterizations; cf. \cite{quartics}---we have
\beq\label{eq:Vgen}
\Delta V 
%&=& m_{H_u}^2 \left| H_u \right|^2 + m_{H_d}^2 \left| H_d \right|^2 \nonumber \\ 
&=&  \lambda_1 \left|H_u^0\right|^4 + \lambda_2 \left|H_d^0\right|^4 -2 \lambda_3 \left|H_u^0\right|^2 \left|H_d^0\right|^2  \nonumber \\
&&\hspace{0cm}+ \left[\lambda_4 \left|H_u^0\right|^2 H_u^0 H_d^0 + \lambda_5 \left|H_d^0\right|^2 H_u^0 H_d^0 \right. \\
&&\hspace{2.85cm} \left.  +\, \lambda_6 (H_u^0 H_d^0)^2  + {\rm c.c.}\right]. \nonumber
\eeq
Note that our conventions differ from others used %commonly employed 
in the literature.  In Eq.~(\ref{eq:Vgen}), the MSSM is recovered by taking
\beq\label{eq:MSSMquartics}
\lambda_1 = \lambda_2 = \lambda_3 = \frac{1}{8} (g^2+g'^2); \ \ \lambda_4 = \lambda_5 = \lambda_6 = 0.
\eeq
As far as the phenomenology of the CP-even states is concerned, the effects of $\lambda_{3,6}$ can be encoded with a single coupling;  we redefine these such that we can consider only $\lambda_3$ in determining the vacuum structure.  

%As in the MSSM, any overall rescaling of the relationships in Eq.~(\ref{eq:MSSMquartics}) will not allow  a suppressed bottom Yukawa when $\tan \beta > 1$.  This leads us to consider the possibility of generating different hierarchies amongst these couplings in an attempt to avoid such a conclusion.  
%In all that follows, we will focus on theories with a softly-broken $Z_2$ symmetry where $\left|\lambda_{1,2,3} \right| \gg \left| \lambda_{4,5}\right|$ such that the latter can be neglected. 
From \Eq{eq:cucd}, we see that suppressing the $h^0$ coupling to bottom quarks requires
$\left| \sin \alpha\right| < \cos \beta,$ or equivalently $|\tan \alpha| < 1/\tan \beta$. For $\tan \beta > 1$, we may translate this into a relatively compact condition on the quartic couplings.  The derivation of this condition is detailed in the Appendix, where we find
\beq
\label{eq:supcondition}
\hspace{-1cm} \lambda_1 \sin^2 \beta  -  \lambda_2 \cos^2 \beta   - \cos (2 \beta) \lambda_3 \nonumber \\  
&&\hspace{-2.3cm} +\, \frac{\sin 3 \beta}{2\cos \beta} \lambda_4 + \frac{ \cos3 \beta}{2\sin \beta} \lambda_5  < 0.
\eeq 
For $\lambda_{4,5}=0$, the condition can be expressed simply:
\beq
(\lambda_1+ \lambda_3) \sin^2 \beta < (\lambda_2 + \lambda_3) \cos^2 \beta.
\eeq 
This illustrates clearly why the quartic structure of the MSSM at tree-level forbids suppression of the down-type Yukawas for $\tan \beta > 1$.

At loop-level in the MSSM, $\tan \beta \gtrsim 5$ is necesary for  a Higgs mass of $125 \, {\rm GeV}$.   Suppression then requires
\beq
\lambda_1 + \lambda_3 - \frac{\lambda_4}{2} \tan \beta \lesssim 0,
\eeq
from which we see a need for corrections that reduce $\lambda_{1,3}$ and enhance $\lambda_4$.  The dominant corrections in the MSSM come from the top sector, which ignoring the logarithms $\ln m_{\tilde{t}}/{m_t}$, are given by \cite{Carena9504316}
%\bea
%\delta \lambda_1 = \frac{3 h_t^4}{8\pi^2} (\bar{A_t}^2-\bar{A_t}^4/12),\quad \delta \lambda_2 = -\frac{h_t^4 \bar{\mu}^4}{64\pi^2},\quad \delta \lambda_3 = \frac{3 h_t^4 \bar{\mu}^2}{64\pi^2} (\bar{A_t}^2-2)
%\delta \lambda_1 = \frac{3 h_t^4}{8\pi^2} (\bar{A_t}^2-\bar{A_t}^4/12),\quad \delta \lambda_3 = \frac{3 h_t^4 \bar{\mu}^2}{64\pi^2} (\bar{A_t}^2-2)
%\eea
\beq \label{eq:toploops}
\delta \lambda_1 = \frac{3 y_t^4}{16\pi^2} (\bar{A_t}^2-\bar{A_t}^4/12), \; &&
\delta  \lambda_3 = \frac{3 y_t^4 \bar{\mu}^2}{64\pi^2} (\bar{A_t}^2-2), \nonumber \\
&& \hspace{-2.3cm} \delta \lambda_4 = \frac{y_t^4 \bar{\mu}}{32 \pi^2} (\bar{A_t}^3-6\bar{A_t}),
\eeq 
where $\bar{\mu} = \mu/m_{\tilde{t}}$ and $\bar{A_t} = A_t/m_{\tilde{t}}$ and definitions of $A$-terms follow those of \cite{Haber}. One first finds that $\delta \lambda_1 > 0$ provided $\bar{A_t}$ is not too large.  In fact, at maximal mixing, where the corrections to the physical Higgs mass are maximized, one requires $\bar{A_t} = \pm \sqrt{6} + \bar{\mu}/\tan\beta.$  Obtaining such a large trilinear term is challenging for many supersymmetry breaking scenarios; this issue and the Higgs mass itself thus suggest that $\delta \lambda_1$ is positive, hurting the required inequality.  The $\delta \lambda_3$ term is positive near maximal mixing, but can be negative for small $\bar{A_t}$.  However, the overall coefficient is small, so one requires $\bar{\mu} \gg \bar{A_t}$ to help satisfy the inequality.  Finally, $\delta \lambda_4$ can be positive for small $\bar{A_t}$ if the sign of $\bar{\mu}$ is opposite that of $\bar{A_t}$.  Interestingly, its size is suppressed near maximal mixing, but can be helped by the $\tan \beta$ enhancement in the inequality.  Through this $\delta \lambda_4$ term, we see why the $\alpha_{\rm eff}$ scenario \cite{Carena0202167, Wagner} requires large $\tan \beta$ and negative $\bar{\mu} \bar{A_t}$ (assuming $\bar{A_t} < \sqrt{6}$) in order to achieve suppressed couplings to the bottom quark. These arguments are not strongly sensitive to the neglected logarithmic corrections $\propto \ln m_{\tilde{t}}/{m_t}$, as the terms in \Eq{eq:toploops} are the matching terms from integrating out the top squarks.  The neglected logarithms are the small corrections coming from renormalization group running down 
%these coefficients 
to the top mass,  which are suppressed by an additional loop factor.  The one important exception is the top quark yukawa contribution to $\delta \lambda_1$, whose  leading logarithmic correction also favors $\delta \lambda_1 > 0$.  

At large $\tan \beta$ there are also significant contributions proportional to $y_b$, but these act identically to the corrections in \Eq{eq:toploops} with $t \to b,$ $\delta \lambda_1 \to \delta \lambda_2$, $\delta \lambda_4 \to \delta \lambda_5;$ in particular the leading correction to $\delta \lambda_3$ from $y_b$ has the same parametric behavior as that from $y_t$.  The bottom correction to the $\tan \beta$ enhanced term $\lambda_4$ is proportional to $\bar{\mu}^3 \bar{A_b}$, so one prefers to have this $\bar{\mu}\bar{A_b}>0$.   For exceptionally large values of $\tan \beta$ ($\tan \beta \gtrsim 50$), there are further corrections due to loop-induced couplings that violate the type-II 2HDM structure, giving a bottom quark mass from $H_u$.  These terms depend on other details of the soft spectrum such as the gluino mass, but can also help in suppressing the bottom coupling.   

%If we assume $\lambda_{4,5}$ to be subdominant, the only way to achieve down-suppression at large $\tan \beta$ is to arrange for a large positive contribution to the coefficient of $\left| H_d^0\right|^4$ or $\left| H_u^0\right|^2 \left| H_d^0\right|^2$ in the potential.  The possibility of realizing the second option, i.e. of establishing $\delta \lambda_3$ large and negative within certain models, will be discussed below.

Even with the full LHC 2012 data set, uncertainties on fermion couplings are projected to be $\mathcal O(50\%)$ \cite{couplings}, so the preference for down-suppression could change.  With this in mind, we note the range of down-type couplings that would be consistent with representative mass ranges of the other Higgs scalars, if such states should be discovered.  Examining $H^0$, for instance, we have scanned the parameter space including stop contributions to the up-type quartic in order to generate $m_h > m_Z$ (we neglect scenarios with large $A$-terms): constraining the light Higgs  to lie in the range $120\, {\rm GeV}\leq m_h \leq 130 \, {\rm GeV}$, we find the coupling ranges
\beq
 m_{H^0} < 300 \; \text{GeV},&& \quad  c_b > 1.3; \nonumber \\ 
300 \; \text{GeV} < m_{H^0} < 500 \; \text{GeV},&& \quad c_b \in [1.1,1.3]; \nonumber \\ 
500 \; \text{GeV} < m_{H^0}, && \quad c_b \in [1,1.1] . \nonumber
\eeq
These conclusions remain consistent at the two-loop level computed using FeynHiggs \cite{FeynHiggs}\footnote{Although the FeynHiggs prediction for $m_h$ in this range of soft masses involves large logarithms that must be resummed, this uncertainty does not significantly impact our results.}.
%With this, we find the results shown in Table~\ref{table:couplings}; these conclusions remain consistent at the two-loop level computed using FeynHiggs \cite{FeynHiggs}\footnote{Although the FeynHiggs prediction for $m_h$ in this range of soft masses involves large logarithms that must be resummed, this uncertainty does not significantly impact the results in Table~\ref{table:couplings}.}.
%\begin{table}[htb]
%\begin{center}
%\begin{tabular}{| c | c |}\hline
%Heavy Higgs [GeV]& $c_d$   \cr   \hline
%$m_{H^0}< 300$ & $ >1.3$  \\ 
%$300< m_{H^0}< 500$  & $\in [1.1,1.3]$  \\ 
%$500< m_{H^0}$ & $\in [1,1.1]$  \\ 
%\hline
%\end{tabular}
%\caption{\small
%Ranges of down-type Yukawas for representative values of the heavy Higgs.  The light Higgs is constrained to the range $120<m_h<130\, {\rm GeV}$, under the assumption that the quantum corrections to its mass arise solely from stops.}
%\label{table:couplings}
%\end{center}
%\label{default}
%\end{table}
Such relationships can help to assess proposals for additional light scalars, for instance as recently discussed in \cite{maiani}. Finally, we note that our analysis has focused on the 2HDM potential and neglected possible loop contributions to Higgs couplings from new light states such as top partners. However as discussed in \cite{Wagner}, corrections from light stops and sbottoms are unlikely to improve the current fit, since the enhancement in the $\gamma\gamma$ branching ratio is counteracted by a suppression in the gluon fusion production of the Higgs.  The conclusion of a suppressed $\gamma\gamma$ rate is generally true of staus and charginos, apart from the exceptional case of maximally-mixed staus \cite{Wagner}.

\subsection{New Dynamics for Down-Suppression}

In order to accommodate suppressed down-like couplings at large $\tan \beta$, we now turn to the introduction of new dynamics to the Higgs potential.  These may arise in the form of additional chiral superfields with couplings to the Higgs or additional gauge fields under which the Higgses are charged. 
%In the latter case, conventional $D$-term corrections \cite{DTerms}  contribute symmetrically to the quartic; 
In the latter case, conventional D-term corrections  \cite{DTerms} contribute symmetrically to the quartic, maintaining the relation $\lambda_1=\lambda_2=\lambda_3$;
generating asymmetric corrections requires the Higgses to be distinguished by the gauge interactions as in \cite{AsymDTerms}. While an appealing possibility, this has extensive ramifications for flavor and we will not pursue such gauge corrections further here. 

The set of possible new matter fields with marginal couplings to the Higgs is constrained by gauge invariance to include only singlets, doublets, and triplets of $SU(2)_W$. Singlets may couple to $H_u H_d$; doublets separately to $H_u, H_d$; and triplets to $H_u H_u$ and $H_d H_d$ as in \cite{triplets}.  The singlet choice is a defining element of the NMSSM and various incarnations (e.g. Fat Higgs, $\lambda$SUSY, etc), where one includes
\beq\label{eqn:Sint}
\Delta W = \lambda S H_u H_d + f(S)
\eeq
where $f(S)$ may include various marginal superpotential terms for $S$ such as tadpole (e.g., Fat Higgs), quadratic (S-MSSM \cite{SMSSM}), or cubic (NMSSM) interactions. The interaction \Eq{eqn:Sint} yields an $F$-term contribution to the Higgs quartic that depends on $\lambda$ and can thus contribute significantly to the Higgs mass. In the case of the NMSSM, $f(S) = \kappa S^3$ and the singlet $S$ acquires a vev. If the soft mass of the singlet is large, doublet-singlet mixing is small and the primary change in the Higgs potential comes from the quartic correction $\delta \lambda_3 = - |\lambda|^2 / 2$ \cite{Blum:2012kn}.  This quartic coupling can flip the sign of $\lambda_3$, making it much easier to satisfy \Eq{eq:supcondition}.  Alternately, we may consider the case when the entire supermultiplet $S$ is massive and may be integrated out above the electroweak scale \cite{DST}. If the singlet acquires a large supersymmetric mass $M_S$, the leading quartic correction is $\delta \lambda_4 = \delta \lambda_5 = - \lambda^2 \frac{\mu^* }{M_S}$ where $\mu$ is an explicit superpotential mass term for $H_u H_d.$  If $\mu^*<0$, these are particularly helpful in \Eq{eq:supcondition}.  If the singlet also acquires a sizable $B \mu$-type nonsupersymmetric mass $\tilde{m}_S$, this results in a correction to the quartic of the form $\delta \lambda_6 = - \lambda^2\frac{\tilde m_S }{2 M_S}$. Finally, when the singlet is relatively light there may be considerable doublet-singlet mixing in the mass eigenstate $h^0$, and it is no longer sufficient to consider the 2HDM potential alone. This mixing suppresses all tree-level couplings of $h^0$ and favors $c_b, c_t < 1$, though it may preferentially suppress the coupling $c_b$ \cite{NMSSMmix}. 

In the case of new hypercharge ($Y=\pm 1$) triplets, couplings may be introduced via interactions of the form
\beq\label{eqn:Tint}
\Delta W = \lambda_T T H_u H_u + \lambda_{\bar T} \bar T H_d H_d + f(T, \bar T) ~.
\eeq
If the scalar triplets are heavy due to a large soft mass, their primary correction to the Higgs potential is of the form $\delta \lambda_1 = |\lambda_T|^2, \delta \lambda_2 =  |\lambda_{\bar T}|^2.$ Alternately, when $T, \bar T$ acquire a large supersymmetric mass $M_T$, the leading quartic correction is of the form $\delta \lambda_4 = \delta \lambda_5 = - 2 \lambda_T \lambda_{\bar T} \frac{\mu^* }{M_S},$ while in the presence of a significant $B \mu$-type soft mass $\tilde{m}_T$ there is a correction $\delta \lambda_6 = - \lambda_T \lambda_{\bar T} \frac{\tilde m_T }{M_T}$ \cite{DST}. 

The final option is to include couplings to new doublets.  In contrast with the singlet and triplet cases, this is done linearly in the superpotential,  i.e.
\beq
\Delta W = \lambda_u H_u \mathcal O_u + \lambda_d H_d \mathcal O_d,
\eeq
where the operators $\mathcal O_{u,d}$ can emerge as composite objects from some new strong dynamics; cf. \cite{SCTC, GP}.  Here the couplings $\lambda_{u,d}$ fix $\beta$, while the (positive) Higgs soft masses can be used to freely tune $\alpha$ along with the Higgs mass itself; the two angles are fully independent in this case. The region of $c_b$-$c_t$ that is closed for the large $\tan \beta$ MSSM is thus reopened, with the possibility of realizing e.g. $c_b \to 0, \ c_t \to 1$ even in the limit $\tan \beta \to \infty$.  The resulting phenomenology could then differ drastically from the MSSM \cite{SCTCpheno}, and would provide  relief if future data single out a region of enhanced up-type couplings.

%%%%%%%%%%%%%%%%%%%%%%%%%%%%%%%%%%%%%%%%%%%%%%%%
%%%%%%%%%%%%%%%%%%%%%%%%%%%%%%%%%%%%%%%%%%%%%%%%
\section{\label{sec:Conclusions} Conclusions}
%%%%%%%%%%%%%%%%%%%%%%%%%%%%%%%%%%%%%%%%%%%%%%%%
%%%%%%%%%%%%%%%%%%%%%%%%%%%%%%%%%%%%%%%%%%%%%%%%
In this paper, we have performed a model-independent fit of current LHC data to the three-dimensional parameter space of tree-level Higgs couplings.
% relevant for supersymmetric extensions of the Standard Model. 
%Such fits may differentiate among supersymmetric theories due to the highly constrained form of the Higgs potential in the MSSM.  
Our fit 
%is consistent with Standard Model Higgs values, but 
shows an interesting preference for deviation from Standard Model Higgs values in the direction of suppressed (enhanced) bottom (top) quark couplings under the assumption that new charged and colored states are sufficiently heavy so as to induce contributions in $h\to gg$ and $h \to \gamma \gamma$ that are relatively small compared to their SM values.   These results are preliminary due to limited statistics, but confirming such nonstandard fermion couplings would have important implications.  Our general type-II 2HDM analysis shows that such couplings are difficult to achieve in the MSSM for $\tan \beta > 1$, due to its tree level Higgs quartic potential, requiring loop corrections pointing to particular corners of supersymmetry breaking parameter space.

As fits improve with increasing statistics, they will provide a lower bound on the mass of additional scalars in the MSSM and may suggest that viable supersymmetric theories of electroweak symmetry breaking involve additional degrees of freedom beyond two Higgs doublets.  To this end, our analysis was used to pinpoint extensions of the supersymmetric Higgs sector that enable bottom suppressed couplings.  In addition, beyond supersymmetry, our conclusions relating the structure of the 2HDM potential to the couplings of the Higgs are very general and may prove useful in interpreting future measurements of Higgs properties. \\

\noindent {\it Acknowledgements---} We thank R.~Contino, M.~Luty, and S.~Thomas for particularly useful discussions on these topics.
NC is supported by NSF grant PHY-0907744, DOE grant DE-FG02-96ER40959, and the Institute for Advanced Study.

\begin{appendix}

\section{Derivation of Condition for Down-Suppression}

To obtain a limit on the quartic couplings, we may express $\tan \alpha$ in terms of the mass-squared matrix for the fluctuations via
\beq
\tan \alpha = \frac{M^2_{uu}-m_h^2}{M^2_{du}} = \frac{-A+\sqrt{A^2+B^2}}{B}
\eeq
where
\beq
A \equiv M^2_{dd}-M^2_{uu}, \quad B\equiv 2 M^2_{du}.
\eeq
These, in turn, may be written as
\beq \label{eq:ab}
A = A_0 - m_A^2 \cos 2\beta, \quad B= B_0 - m_A^2 \sin 2\beta
\eeq
where $A_0, B_0$ have no $m_A$ dependence and take the form
\beq \nonumber
A_0/v^2 &=& \left[2 \lambda_2 \cos^2 \beta - 2 \lambda_1 \sin^2 \beta -\frac{\lambda_4}{2}\left(\sin 2\beta + \tan \beta\right) \right. \\
&&\left. +\frac{\lambda_5}{2}\left(\sin 2\beta + \cot \beta\right)\right]\\
B_0/v^2 &=& - 4\lambda_3 \sin \beta \cos \beta + 3 \lambda_4 \sin^2 \beta + 3\lambda_5 \cos^2 \beta
\eeq

We can see from Eq.~(\ref{eq:ab}) that in the limit of large $m_A$, $\tan \alpha_\infty = -1/\tan \beta$, corresponding to the usual decoupling limit,  $\alpha = \beta-\pi/2$.  There are no extrema in $\tan \alpha$ in the relevant range of $m_A^2 \in [0,\infty]$, but there may be discontinuities if $B$ vanishes.  Since $B_\infty < 0$, $B$ will cross zero only if $B_0>0$.  This crossing point occurs when $m_A^2 = B_0/\sin 2\beta$.  At this point, $A_{cross} = A_0 -B_0 \cot 2\beta$.  If $A_{cross}> 0$, then $\tan \alpha_{cross} = 0$ and by continuity there will be a region of suppressed coupling to bottom quarks.  However, if $A_{cross} < 0$, $\tan \alpha = \text{sign}(B) \times \infty$ at the crossing.    Thus, by continuity, the requirements to have a region of suppressed bottom quark coupling ($|\tan \alpha| < 1/\tan \beta$), are 
\begin{enumerate}
\item  If $B_0<0$, then  $\tan \alpha_0 > -1/\tan \beta$
\item  If $B_0>0$ and $A_{cross} < 0$ then $\tan \alpha_0 < 1/\tan \beta$
\item  If $B_0>0$ and $A_{cross} > 0$ there is always a suppressed region
\end{enumerate}

Case 1 occurs when $\tan \alpha_0 > - 1/\tan \beta;$   this requires
\beq
\left[-\frac{A_0}{|B_0|} + \sqrt{\left(\frac{A_0}{|B_0|}\right)^2+1}\right] < 1/\tan \beta \\ \nonumber
 \text{or} \quad \quad \quad \quad \quad \quad\\
\frac{A_0}{|B_0|} >  \frac{1}{2}(\tan\beta - \cot \beta) . \quad \quad \quad \quad
\eeq

Case 2 is never satisfied when $\tan \beta >1$.  To see this, assume $B_0 >0$, in which case $A_{cross} = A_0 - B_0 \cot 2\beta$.  In Case 2 we also have $A_{cross} < 0$, which requires that $A_0/|B_0|  < \cot 2 \beta$.  However, $\tan \alpha_0 < 1/\tan \beta$ requires that $A_0/|B_0| > \frac{1}{2}(\tan\beta - \cot \beta).$  These are compatible inequalities if $\frac{1}{2}(\tan\beta - \cot \beta)< \cot 2\beta.$  This compatibility only occurs for $\tan \beta < 1$.  Thus, Case 2 is impossible to realize for $\tan \beta > 1$.

Case 3 occurs when $B_0 > 0$ and $A_0/|B_0| > \cot 2 \beta = -\frac{1}{2} (\tan \beta - \cot \beta)$.  Thus combined with Case 1, suppression of the bottom quark coupling for $\tan \beta > 1$ requires
\beq
A_0 > -\frac{B_0}{2}(\tan\beta - \cot \beta)
\eeq  
which translates into the following condition on the quartic couplings:
\beq
\hspace{-1cm} \lambda_1 \sin^2 \beta  -  \lambda_2 \cos^2 \beta   - \cos (2 \beta) \lambda_3 \nonumber \\  
&&\hspace{-2.3cm} +\, \frac{\sin 3 \beta}{2\cos \beta} \lambda_4 + \frac{ \cos3 \beta}{2\sin \beta} \lambda_5  < 0.
\eeq 
This general condition agrees with the leading term as one takes the decoupling limit, as explored in \cite{Gunion:2002zf, Randall:2007as,Blum:2012kn}.
%This constrains $\lambda_3$ to lie in the range
%\beq
%\lambda_3 \in \left[\min(A,B),\max(A,B) \right]
%\eeq
%where
%\beq
%A&=& -\frac{B\mu}{v_u v_d} + \frac{\lambda_1 v_u^2 - \lambda_2 v_d^2}{v_u^2 -v_d^2} \nonumber \\
%&&+\, \frac{\lambda_4 v_u^3}{v_d \times (v_u^2 -v_d^2)}-\frac{\lambda_5 v_d^3}{v_u \times (v_u^2 -v_d^2)};\\
%B&=& -\frac{\lambda_1 v_u^2 + \lambda_2 v_d^2}{v_u^2 -v_d^2} \nonumber \\
%&& + \frac{\lambda_4 v_u (v_u^2 - 3 v_d^2)}{2 v_d (v_u^2 -v_d^2)}- \frac{\lambda_5 v_d (v_d^2 - 3 v_u^2)}{2 v_u (v_u^2 -v_d^2)}.
%A &=& \frac{\lambda_2 v_d^2 -\lambda_1 v_u^2}{v_u^2 -v_d^2} \\
%B &=& -\left(\frac{B\mu}{v_u v_d} +  \frac{\lambda_2 v_d^2 -\lambda_1 v_u^2}{v_u^2 -v_d^2}\right).
%\eeq

%In the general 2HDM, $\alpha$ can be positive; in this case the condition for down-suppression at $\tan \beta >1$ becomes
%\beq
%\frac{v_u^2 \times (\lambda_1 - \lambda_3)}{v_u^2 -v_d^2} -\frac{v_d^2 \times (\lambda_2-\lambda_3)}{v_u^2 -v_d^2} > \frac{B\mu}{v_u v_d} .
%\eeq
%In cases of simple deviations from the MSSM, Eq.~(\ref{eq:Inequality}) is the sufficient condition for down-suppression.

%We see from Eq.~(\ref{eq:Inequality}) why the Yukawa couplings of the MSSM can't access the full space of the 2HDM at large $\tan \beta$.  Note in particular that loop contributions from the stop to the up-type quartic (which crucially serve to increase $\lambda_1$) work against us when trying to suppress the bottom coupling. 

\end{appendix}

\end{document}